# Nonlinear nano-electromechanical lattices for high-frequency, tunable stress propagation

**Authors:** Jinwoong Cha[1,2], Chiara Daraio[*,2]

**Affiliations:**

[1]Department of Mechanical and Process Engineering, ETH Zurich, Switzerland

[2]Engineering and Applied Science, California Institute of Technology, Pasadena, CA, USA

*Correspondence to: Prof. Chiara Daraio (daraio@caltech.edu)

**Active manipulation of mechanical waves at high frequencies opens opportunities in heat management[1,2], radio-frequency (RF) signal processing[3-5], and quantum technologies[6-10]. Nanoelectromechanical systems (NEMS) are appropriate platforms for developing these technologies, offering energy transducibility between different physical domains, for example, converting optical or electrical signals into mechanical vibrations and vice-versa. Existing NEMS platforms, however, are mostly linear, passive, and not dynamically controllable. Here, we report the realization of active manipulation of frequency band dispersion in one-dimensional (1D) nonlinear nanoelectromechanical lattices (NEML) in the RF domain (10-30 MHz). Our NEML is comprised of a periodic arrangement of mechanically coupled free-standing nano-membranes, with circular clamped boundaries. This design forms a flexural phononic crystals with a well-defined band gaps, 1.8 MHz wide. The application a DC gate voltage creates voltage-dependent on-site potentials, which can significantly shift the frequency bands of the device. Dynamic modulation of**

**the voltage triggers nonlinear effects, which induce the formation of phononic band gaps in the acoustic branch. These devices could be used in tunable filters, ultrasonic delay lines and transducers for implantable medical devices.**

Architectured materials, such as photonic, plasmonic, and phononic metamaterials, control wave propagation with specifically designed geometry of their building blocks. For application in devices, it is desirable to be able to tune their operational frequencies on demand. Effective tuning mechanisms exploit mechanical compliance to alter the architecture geometry, including structure and/or periodicity[11-15], but this approach is limited to the use of soft materials. Other tuning mechanisms involve the change of intrinsic properties of constituent materials, e.g., in photonic crystal this is achieved by changing the dielectric constant with heating[16] and in plasmonic metamaterials by changing the charge carrier density or fermi level of materials[17-20]. Phononic metamaterials present an advantage over their optical and electronic counterparts, in that they can more easily access nonlinear regimes for tunability. In the small amplitude domain, elements in phononic crystals can be strongly coupled, enabling controlled energy transfer in space, like in phonon waveguides[21,22]. At larger amplitudes, mechanical deformations can access nonlinear regimes that allow energy transfer across different energy states, or vibration modes, allowing, for example, phonon cavity dynamics[23-25] and synchronization[26]. Macroscopic closed-packed granular crystals exploit the nonlinear contact mechanics between two spheres, to change the effective coupling strengths. The application of a small static pre-compression, can be used to demonstrate tunable band gaps[27], wave velocity[28], and wave localization[29]. In other examples, magnetic repulsive forces have been used to control the on-site potential of locally resonant units, thereby realizing phononic transistors[30] and programmable metamaterials[31]. However, due to scaling constraints such tunability could not be demonstrated at device relevant frequencies, in the ultrasonic domain.

Nanoelectromechanical system (NEMS) with voltage induced frequency tuning[32,33] offer a practical platform to miniaturize phononic devices, making megahertz and gigahertz transduction accessible with conventional fabrication methods. The application of a DC voltage to a single resonator demonstrated the ability to trigger different nonlinearities, depending on the thickness to vacuum gap ratio. The DC voltage increased the resonant frequency by tension-induced stiffening[32,33] or decreased it by electrostatic softening of the resonators[34] (Supplementary Fig. S1). Two or three coupled-NEMSs with such tunability have shown interesting nonlinear dynamic phenomena including cooling and amplification[25]. Recently, NEMS based phononic crystals consisting of suspended GaAs/AlGaAs heterostructures[21,22] have been suggested as RF waveguides. They demonstrated Bragg phononic band gaps, slow phonons, and defect-mediated nonlinear dynamics. Although promising, these devices are still non tunable, meaning that the phononic properties are pre-determined during fabrication. In addition, the defect-mediated dynamic control has a narrow bandwidth and low modulation efficiency, transmitting substantial energy to the stop band. In this study, we realize a tunable phononic crystal that harnesses DC/AC electrostatic forces exerted on the membranes, to control wave transmission over a broad frequency range.

Our NEML consists of Si-rich LPCVD silicon nitride ($SiN_x$) membranes (10 $\mu$m in diameter) that form a phononic crystal with periodic, curved boundaries (Fig. 1a). $SiN_x$, an insulator, is used to electrically separate the excitation and the tuning electrodes. Neighboring membranes are overlapped to create mechanical coupling (Fig. 1b and 1c). We construct membrane arrays with variable numbers of repeating units (Supplementary Fig. S2). The periodicity of the lattice is chosen to be $a = 7$ $\mu$m owing to the expected frequency dispersion, with prominent acoustic and optical branches separated by a well-defined phononic band gap (Fig. 1d). To characterize the devices, we measure the frequency response of NEML with and without tuning electrodes using a laser interferometer (see Methods). Flexural motion of the membranes is triggered by simultaneously applying DC and AC voltages (Fig. 1a). As shown in Figure 1d, the spectra of NEML without and with electrodes present clear stop bands below 12 MHz and 11.5 MHz, respectively. The NEML without electrodes has a band gap from 17 MHz to 19.8 MHz, while the NEML with electrodes has a gap from 16 MHz to 17.8 MHz, confirming that the gold electrodes do not significantly change the operating frequencies. Passband spectra show discrete peaks due to the finite number of unit cells (N = 120). We perform frequency response measurements with varying number of unit cells (N = 30, 60, 90, and 120) and observed well-defined phononic band gaps even at the lowest unit cell numbers (Supplementary Fig. S3). The average quality factor (Q-factor) of the device with (without) electrodes is 1700 (2800) and it depends on the surface and the intrinsic material loss due to the evaporated gold[35]. For applications in quantum state transfer that requires low loss, the Q-factor can be enhanced by replacing the gold electrodes with graphene[36], increasing the residual stress of the film, and operating in cryogenic environments.

To inform our experiments, we perform numerical simulations using Comsol Multiphysics© with Floquet periodic boundary condition, considering geometric nonlinearity. The experimental dispersion curves, obtained from fast-Fourier transformation of the spatially scanned data, capture the phononic behavior of our NEML (Supplementary Fig. S4). The experimental results show good agreement with numerical simulations (Fig. 2a-d) at gating voltages ($V_T$) ranging from 0 V to 22 V, capturing correctly the location and the size of pass and stop bands. Flat band regions exist near the band edges, at $q$=0 and $\pi/a$, suggesting slow, near zero group velocities. The locally clamped boundaries, acting as grounding springs, cause the acoustic band edge at $q$=0 to start from finite frequencies. To systematically investigate the dependence of the transmission properties on the applied gate voltage, we measure the frequency response of the last resonator (located on the opposite boundary from the excitation) while varying the gate voltage from 0 V to 25 V (Fig. 2f). We note a broadening of the acoustic branch from 4.5 MHz at $V_T = 0$ V to 6 MHz at $V_T=25$ V and a decreases of the band gap size from 1.8 MHz at $V_T = 0$ V to 1.3 MHz at $V_T = 25$ V.

To explain the experimental observations, we developed an analytical reduced order model of the finite samples based on discrete interconnected rigid plates (Supplementary Fig. S5). This model clarifies the interplay of the experimental parameters and their effect on the dispersion response of the NEML (Supplementary Fig. S6 and S7). The presence of a gate voltage below the membrane acts as an on-site potential that tunes the transmission. An increase of the electrostatic force applied by the tuning electrodes softens the on-site potentials and consequently down-shifts the frequency bands. In addition, the presence of clamped boundaries causes the development of an axial tension along the lattice. In the discrete model, this tension decreases the contrast of the rotational couplings, reducing band gap size. The boundary effect is not captured by the finite element model that assumes periodic boundary conditions (see Methods). The experimentally measured variations of the center frequency and the band size match well the numerical data (Fig. 2f and 2h) below $V_T = 23$ V. We characterize the Q-factor as a function of the gating voltage of the NEML and show that it decreases quadratically due to dielectric losses (Fig. 2i).

The tunability of the dispersion response means that the transmissible frequencies can be dynamically selected, but also that the group velocity can be controlled. This is evident from the transient responses of the lattice at $V_T = 0$ V and 24 V (Fig. 3a and 3b). We send a chirped pulse with frequencies between 8 MHz ~ 22 MHz through the lattice, and measure the responses of the end unit. The measured response contains several displacement pulses, which originate from boundary reflections. To analyze the dispersive behaviors at single frequencies, we apply a Burtterworth filter with 200 kHz bandwidth to the broadband responses. We then plot the envelope of the filtered pulses in time domain, at each frequency (Fig. 3d and 3f). The time-of-flight of the wave near the band edges is much longer than that in the middle of pass bands, confirming the strong dispersion near the band edges. The experimental group velocity is obtained calculating the velocity of the traveling pulses ($v_s$) at each frequency, using $v_s = 2Na/\Delta t$ (Figure 3c and 3e). Here, $N=120$ is the number of unit cells, $a$ is the periodicity and $\Delta t$ is the time-of-flight. The factor 2 accounts for reflections. The experimental group velocities show excellent agreement with the numerical group velocities. The small discrepancy found in the group velocities at $V_T = 24$ V (Fig. 3e) originates from boundary effects. The envelope of waves near the acoustic band edge (14.5 MHz) broadens due to dispersion. Propagation losses are deduced from the amplitude decay of two neighboring pulses at 13.5 MHz and are found to be around 25% in voltages or 44% in energy per 1.68 mm travelled, which is equivalent to 2.6 dB/mm. At 12 MHz, the pulse speed at $V_T = 24$ V is about twice than the pulse speed at $V_T = 0$ V, demonstrating group velocity tuning. This observation is important for signal processing applications, e.g., in tunable delays line.

Metamaterials under dynamic environments have shown many interesting features, e.g., non-reciprocity[37] and non-equilibrium phenomena[38]. We investigate the behavior of our system under dynamic, AC tuning. In addition to the AC voltage, we apply a 5V DC voltage to the tuning electrode to increase the modulation amplitude. This DC voltage is small enough not to induce significant softening effects. Since the tuning electrode simultaneously excites the entire lattice, we expect their modulation to predominantly affect the lowest modes in the phononic band. We note that the AC tuning triggers nonlinear dynamic phenomena in the NEML leading to a classical hysteretic behavior (Fig. 4a). The resonance curves show a hardening behavior, due to tension induced nonlinearity. Beyond the bifurcation point, at modulation frequencies ($\omega_p$) above 11.3 MHz, a small band gap emerges in the acoustic branches (Fig. 4b and 4c). The gap size increases with increasing modulation frequency, changing its position. The shift of band gap originates from the up-shift of the entire dispersion curves, due to stronger tension induced at higher modulation frequencies. The presence of the modulation-induced band gap is also evident in the real space scanning data obtained at $\omega_p$=11.5 MHz, with $2a$ = 14 μm scanning step (Fig. 4d). Analysis of the corresponding dispersion curve confirms the effective "dimerization" of the unit cell, with a gap at the new band edge $q=\pi/2a$ (Fig. 4e), at a center frequency around 12.7 MHz. Accordingly, the nonlinear second harmonics of the modulation frequency lies at $q=\pi/2a$ in the dispersion relation and the modes' amplitudes at $q=\pi/2a$ have $2a$ periodicity (Supplementary Fig. S8 to S10).

Here, we demonstrate the static and dynamic control of nanoelectromechanical lattices by electrostatic forces. Their frequency dispersion can be tuned over a wide range by applying DC gate voltage. The DC tuning can be used to shift the band edges, reduce the width of the band gap, and control the group velocity. Dynamic AC modulation of the gate shows a new mechanism for band gap formation, induced by nonlinear resonances. This NEML contributes to the development of miniaturized high-frequency components, such as tunable filters and delay lines for signal processing applications, with smaller foot-print compared to the electromagnetic counterparts.

**Acknowledgements**

We acknowledge partial support for this project from NSF EFRI Award No. 1741565. We thank Emre Togan at ETH Zurich for his advice on interferometers.

**Author contributions**

J.C. and C.D. conceived the idea of the research. J.C. designed and fabricated the samples. J.C. built the experimental setups and performed the experiments. J.C. developed the analytical models and performed the numerical simulations. J.C. and C.D. wrote the manuscript.

**Competing Financial Interests**

Nothing to report.

**Methods**

**Sample fabrication**

The fabrication of the nanoelectromechanical lattices begins with low pressure chemical vapor deposition (LPCVD) of 100 nm low-stressed Si-rich silicon nitride ($SiN_x$) on both sides of a $SiO_2$(150 nm)/Si(525 $\mu$m) wafer. The sacrificial thermal oxide under the $SiN_x$ layer defines the vacuum gap distance when the device is released in the HF etching step. The wafer is then cleaned with acetone and iso-proylalchol (IPA). We then spin-coat 495K A4 and 950K A2 Poly-methly-methacrylate (PMMA). The electrode patterns are transferred using an electron-beam (e-beam) lithography, and the PMMA resists are developed in a MIBK and IPA mixture. 45 nm-gold and 5 nm-chromium layers for e-beam alignment markers, excitation and tuning electrodes are then deposited using an e-beam evaporator. Then, the samples are immersed in acetone for 2 hours to lift of the remaining PMMA resists covered by the metals. For the next e-beam exposure, we spin-coat 500 nm ZEP520 e-beam resist to serve as an etch mask for the following reactive ion etching process. A second e-beam lithography is then performed to pattern the arrays of the etch holes with 500 nm diameter and the exposed resist is developed in a ZED-N50 developer. We employ an inductively coupled plasma (ICP) dry etching to drill the holes in the $SiN_x$ device layer. The remaining resist is dissolved in a N-Methyl-2-pyrrolidone (NMP) based solvent. The wafer is then scribed into 3 mm × 3 mm dies and the dies are immersed in Buffered Oxide (BOE) solution to etch the sacrificial $SiO_2$ layer through the etch holes. After the etching process, we use a critical point dryer to avoid the adhesion of the released free-standing membranes to the substrates and obtain perfectly isotropic circular membranes. The Si-rich SiN$_x$ is hardly etched by buffered HF (etch rate of ~0.3 nm/min). Furthermore, the LPCVD process leads to minimal disorder and structural defects, ensuring uniform adhesion to the sacrificial layer, and low residual stress (~100 MPa).

**Experiments**

The measurements of the mechanical motion of the membranes are performed using a home-built optical interferometer, at room temperature and a vacuum pressure P < 10$^{-6}$ mbar. The interferometer is a phase-locked Michelson interferometer that employs a balanced homodyne detection scheme. The phase-lock is enabled by moving a reference mirror mounted on a PID-controlled piezoelectric actuator at 1.5 kHz. The motion of the membranes is electrostatically excited by simultaneously applying DC and AC voltages through a bias tee (Mini-circuits, ZFBT-6GW+). A 633 nm laser light with 20 $\mu$W input power is incident on the sample. The reflected light interferes with the light from the reference mirror. The intensity of the interfered light is measured using the balanced photodetector. Its RF output signal is connected to high frequency lock-in amplifier (Zurich instrument, UHFLI). The position of the laser spot was monitored with a CMOS camera and the measurement position was controlled by a computer-controlled XY-linear stage supporting the vacuum chamber. Details of the method can be found in the Supplementary Information.

**Numerical Simulations**

All the numerical simulations to calculate the phononic frequency dispersion are performed using the finite element method (FEM) via COMSOL. We employ the module pre-stressed eigenfrequency analysis considering geometric nonlinearities, to reflect the effect of membrane deflection and stresses induced by the DC gate voltage. The DC voltage also introduces an effective softening potential and we implement this effect by assigning a negative elastic foundation. To model the elastic foundation, we use an effective gap distance $d_{\text{eff}}$ = 180 nm (Supplementary Note 3). The parameters used for silicon nitride films are 3000 kg/m$^3$ density, 290 GPa Young's modulus, and 0.27 Poisson ratio. For the properties of the electrodes made of 45 nm gold and 5 nm chrome, we calculate the geometric averages of the properties and obtain 18100 kg/m$^3$ density, 98 GPa Young's modulus, and 0.417 Poisson ratio. We impose anisotropic in-plane residual stresses 35 MPa for the direction perpendicular to the lattice and 125 MPa for the parallel direction, to match the experimental dispersion. The contributions of both residual stress components are discussed in the Supplementary Information.

38. Bachelard, N., *et al.* Emergence of an enslaved phononic bandgap in a non-equilibrium pseudo-crystal, *Nat. Mat.* **16**, 808-813 (2017)

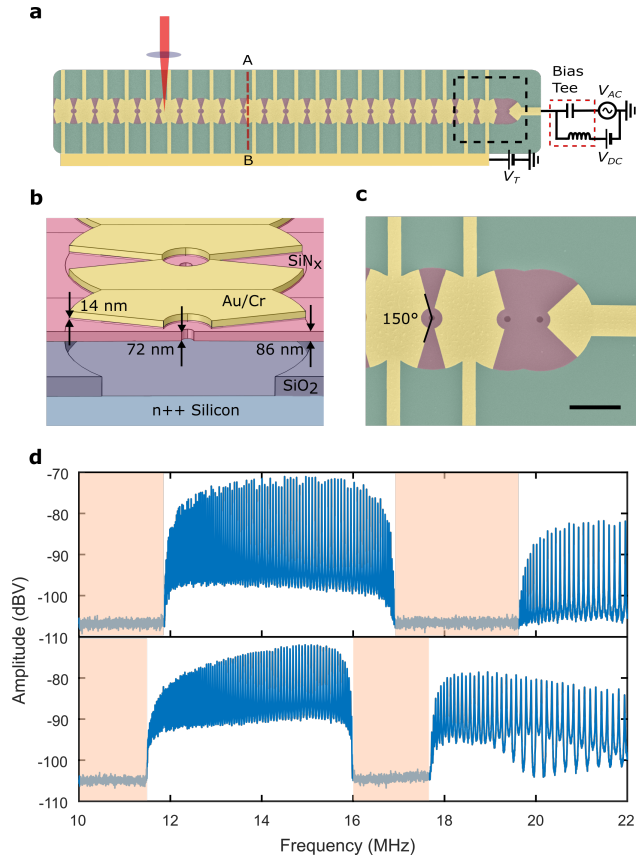

**Figure 1. Nonlinear nanoelectromechanical lattice.** (**a**) Scanning electron microscope (SEM) image of the nonlinear nanoelectromechanical lattice. Flexural motions are electrostatically triggered by simultaneously applying DC ($V_{DC}$) and AC ($V_{AC}$) voltages to the excitation electrode via a bias-tee. $V_T$ is a tuning gate voltage applied to the tuning electrodes to activate on-site electrostatic potentials. The membrane motions at desired points are measured using a home-built balanced homodyne interferometer. (**b**) Schematic of cross-section A-B of the device marked in (**a**). The thickness of the metal electrodes and the sacrificial oxide are 50 nm and 150 nm, respectively. A highly doped n-type silicon substrate is used as a electrical ground plane. (**c**) A magnified SEM image of the dashed-square box in (**a**) shows the region near the excitation electrodes. Scale bar, 5 $\mu$m. (**d**) Frequency response of the lattices with 120 unit cells. The upper (lower) panel presents the response of the device without (with) electrodes. The shaded area indicates frequency stop bands.

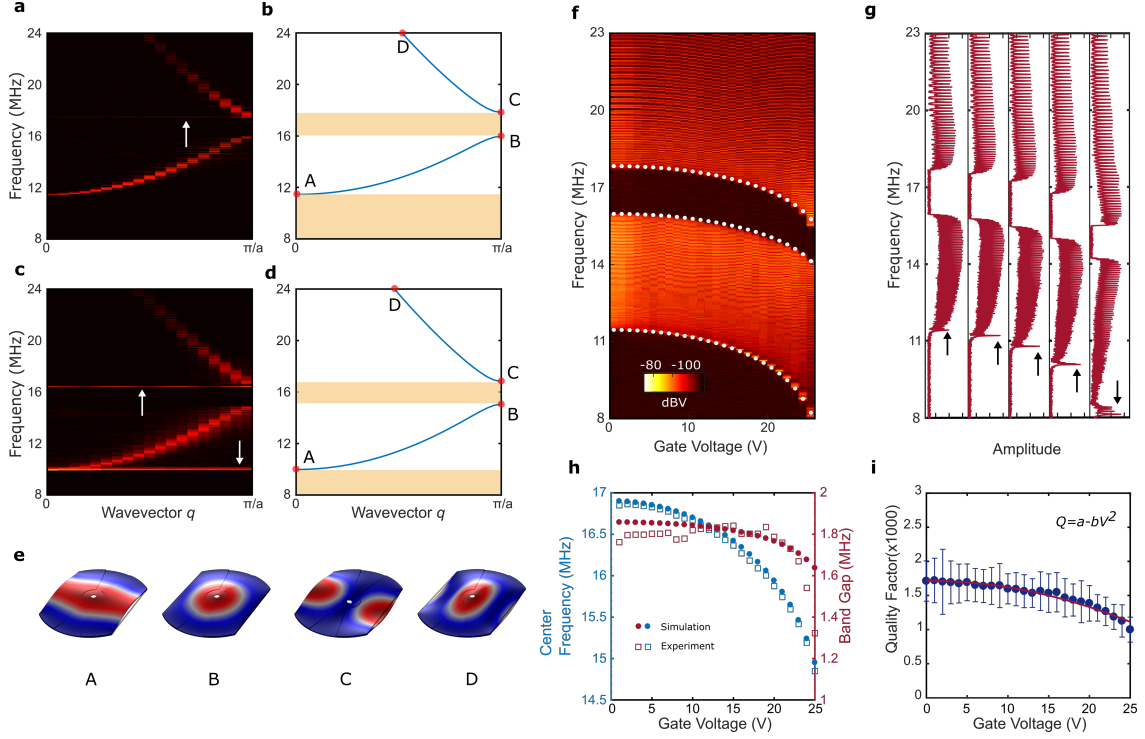

**Figure 2. Static tuning with a DC gate voltage.** (**a**), (**c**) Experimental frequency dispersion curves for $V_{DC}$ = 2 V, $V_{AC}$ = 40 mV, (**a**) $V_T$ = 0 V, and (**c**) $V_T$ = 20 V. The two upward poiting arrows in (**a**) and (**c**) indicate the defect mode of the excitation unit in the front-end of the lattice. The bottom white arrow in (**c**) indicates several modes excited by parasitic couplings between the excitation and tuning electrodes. (**b**), (**d**) Numerical frequency dispersion curves corresponding to (**a**) and (**c**). The light yellow regions indicate the stop bands. (**e**) Mode shapes of a unit cell at different frequencies A, B, C, and D marked in (**b**) and (**d**). (**f**) Experimental frequency responses of the NEML at different applied gate voltages ranging from $V_T$ = 0 V to 25 V. Here, $V_{DC}V_{AC}$ = 0.08 V² is kept constant. The white dots denote the corresponding numerical band edges at $q$ = 0 and $q$ = $\pi$/a. (**g**) Frequency responses varying $V_T$ from 5 V to 25 V with 5 V step (left to right). The black arrows indicate the modes that stem from the parasitic coupling between the excitation and tuning electrodes. (**h**) Position of the band gap center and size as a function of $V_T$. (**i**) Averaged quality-factor from the frequency responses as a function of $V_T$. The quality-factor shows $Q \sim -V^2$ scaling with fitting parameters $a$ = 1717, $b$ = 0.9670.

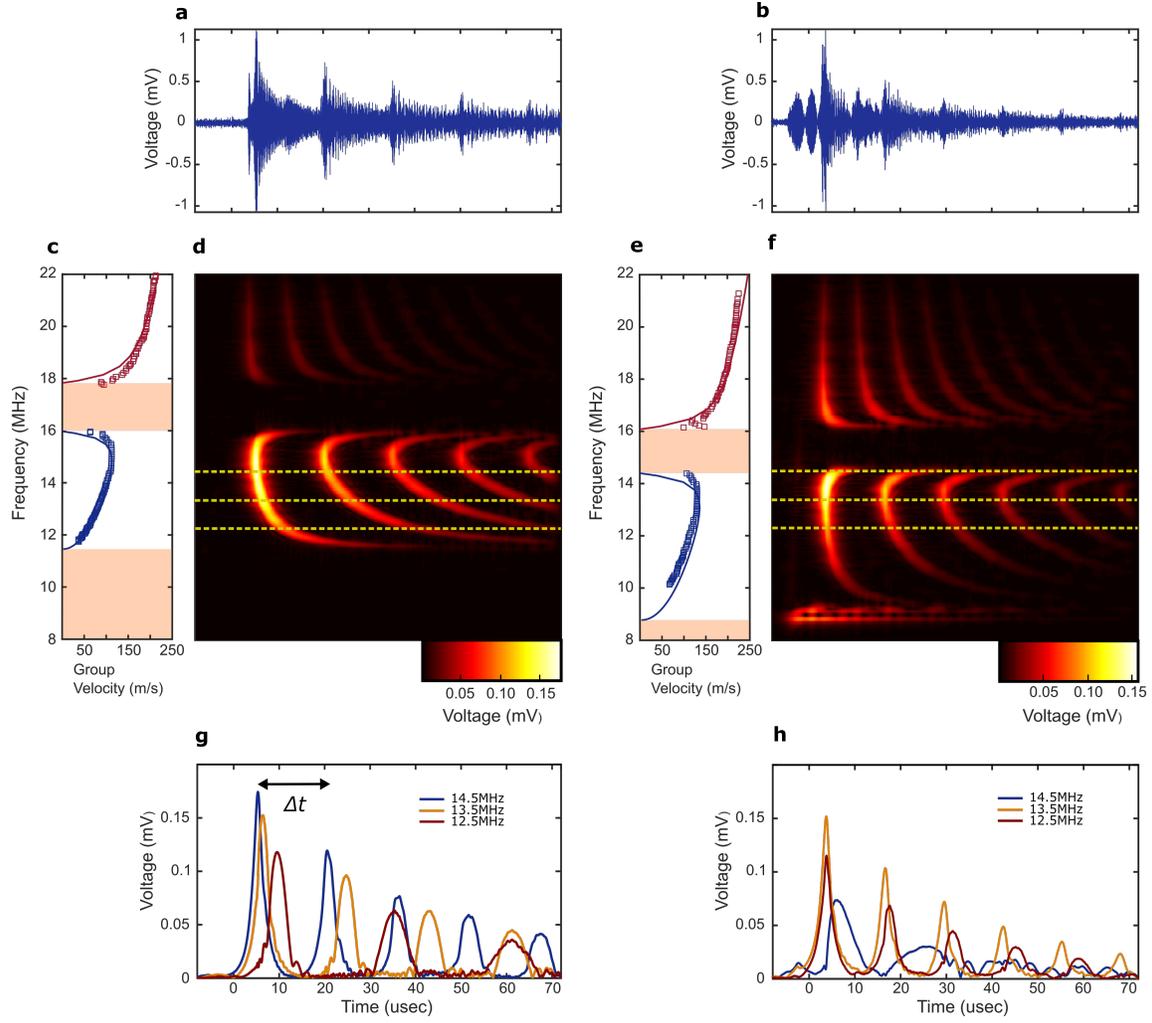

**Figure 3. Tunable phonon propagation velocity.** (**a**), (**b**) Transient responses of the displacement of the end resonator for (**a**) $V_T = 0$ V and (**b**) 24 V prior to the signal filtering process. A chirped signal with frequency content ranging from 8 MHz to 22 MHz is applied to the excitation electrode with $V_{DC} = 20$ V, $V_{chirp,peak} = 375$ mV. (**c**), (**e**) Group velocity as a function of the center frequency of an enveloped pulse for (**c**) $V_T = 0$ V and (**e**) 24 V. The blue and red solid lines in (**c**) and (**e**) represent the group velocities for the acoustic and optical branches obtained from numerical simulations. The dotted open squares are the corresponding experimental group velocities obtained by calculating the time of flight of a single round-trip of a single pulse at different frequencies. (**d**), (**f**) The color maps represent the time-domain amplitude responses of signal envelopes at different frequencies for (**a**) $V_T = 0$ V and (**b**) 24 V. A Butterworth filter at single frequencies with 200 kHz bandwidth is applied to filter out the broadband transient signals in (**a**) and (**b**). Due to the increased AC parasitic coupling, several lowest modes are excited around 9 MHz, as shown in (**f**). (**g**), (**h**) Time-domain pulse envelopes at 12.5 MHz, 13.5 MHz, and 14.5 MHz, corresponding to the three yellow-dashed lines in figure (**d**) and (**f**). Δt is the time of flight of an envelope.

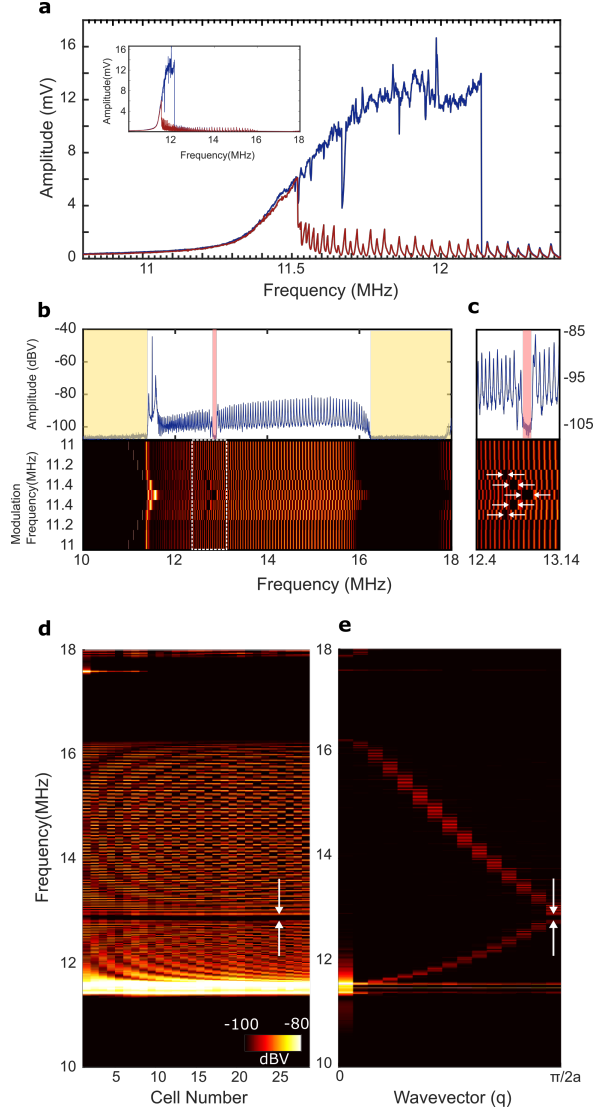

**Figure 4. Dynamic modulation and formation of nonlinear band gap.** (**a**) Frequency spectrum obtained when the tuning electrodes are nonlinearly driven at $V_{DC} = V_{T,DC} = 5$ V, $V_{AC} = 0$ mV, and $V_{T,AC} = 100$ mV. The blue (red) solid line denotes a forward (backward) frequency sweep. The modes in the pass bands can be excited due to the AC coupling with the excitation electrode. The inset shows the broad band frequency spectra ranging from 10 MHz to 18 MHz. (**b**) The top panel shows the frequency response of the lattice subject to a dynamic modulation with a modulation frequency $\omega_\pi$=11.5 MHz, with $V_{DC} = V_{T,DC} = 5$ V, $V_{AC} = 10$ mV, $V_{T,AC} = 100$ mV. The yellow shaded regions are the original band gap and the pink region indicates the dynamically induced band gaps. The color map shows the frequency responses at different modulation frequencies ranging from 11 MHz to 11.5 MHz, with 1 MHz step. We perform a forward and backward sweep of the modulation frequency to check the hysteresis. (**c**) The zoomed-in data in the white dashed box in (**b**) and the corresponding frequency responses. (**d**) Spatially scanned data along the 28 sites in the lattice. The induced band gap is highlighted by two white arrows. The bright regions at low frequency originate from the strong dynamic modulation. (**e**) Frequency dispersion curves corresponding to the scanned data in (**d**). White arrows at $q = \pi/2a$ indicate the band gap.

16